\documentclass[sigchi, nonacm]{acmart}

\usepackage{booktabs}
\usepackage{graphicx}
\usepackage{xcolor}
\usepackage{enumitem}
\usepackage{subcaption}
\usepackage{balance}
\usepackage{multirow}
\usepackage{amsmath}
\usepackage{tabularx}
\usepackage{threeparttable}
\usepackage{array}

\newcommand{\system}{\textsc{Red-Rec}}
\newcommand{\n}{\textit{n}}

\title{From Passive Feeds to Guided Discovery: AI-Initiated Interaction for Vague Intent in Content Exploration}

\author{Yu Xie}
\email{xieyu6@xiaohongshu.com}

\author{Ying Qi}
\email{yingqi1@xiaohongshu.com}

\begin{document}

\begin{abstract}
Recommendation feeds work well when people are simply browsing, and search works well when they can formulate a query. Between these two cases is a common but poorly supported state: users feel that their feed has become repetitive, yet cannot clearly specify what they want instead. We refer to this state as \textit{vague intent}. We present \system{}, an AI-supported exploration interface for this middle ground. After a period of browsing, the system summarizes patterns in the current feed (e.g., dominant content categories and possible latent interests), offers clickable exploration options, asks at most one follow-up question, and then gradually blends new content into the feed. The design is motivated by a formative study (\n{}=12), which found that users often recognize feed staleness but struggle to articulate alternatives, suggesting the need for proactive and low-effort interaction.We evaluated \system{} in a mixed-design lab study (\n{}=28) against three comparison conditions: a passive feed, search, and a user-initiated chat interface. Compared with user-initiated chat, \system{} led to broader exploration ($M$=7.4 vs.\ 4.1 categories, $p<.001$), higher serendipity ratings ($M$=5.72 vs.\ 4.31, $p<.01$), and lower interaction effort. Participants in the AI-initiated condition typed very little (median = 0 characters), relying mainly on option selection, whereas participants in the user-initiated chat condition typed substantially more (median = 73 characters). We discuss how proactive, option-based AI support can help users move beyond repetitive feeds without undermining their sense of control, and we outline design implications for recommendation interfaces that support open-ended exploration.
\end{abstract}



\maketitle

\section{Introduction}

Recommendation feeds and search interfaces support two different modes of content discovery. In one mode, users browse without a specific goal and rely on the system to decide what to show next. In the other, users have a clear information need and express it through a query. Both interaction patterns are widely used and well supported in current platforms.

In our early interviews, however, participants repeatedly described a state that fit neither mode. They felt that their feed had become repetitive or stale, and they wanted the content to shift in some direction, but they could not clearly say toward what. We use the term \textit{vague intent} to describe this situation: the user has a desire to redirect or broaden what they see, but does not yet have a target topic, query, or filter adjustment in mind. people say "my feed is boring" "I keep seeing the same stuff" or "I want something different but I don't know what" The recommendation algorithm is doing exactly what it was trained to do---maximize engagement on historical preferences---and the result, over weeks and months, is a feed that converges on a narrow slice of content~\cite{pariser2011filter, nguyen2014exploring, bakshy2015exposure}. Users feel it. They just cannot do anything about it.

The reason they cannot is an interface gap. Today's content discovery tools serve two extremes of a demand spectrum (Table~\ref{tab:spectrum}). Recommendation feeds handle the case where you have no intent at all---you scroll, the algorithm decides. Search handles the case where you know what you want---you type a query, the system retrieves. Both are well-optimized for their respective jobs.

\begin{table}[t]
\centering
\caption{The demand spectrum for content discovery. The middle segment---vague intent---is prevalent but lacks dedicated interaction support.}
\label{tab:spectrum}
\small
\begin{tabularx}{0.5\textwidth}{l l X}
\toprule
\textbf{Intent Type} & \textbf{Current Interface} & \textbf{User State} \\
\midrule
No intent & Recommendation feed & "Just browsing" \\
\textbf{Vague intent} & \textbf{(Unserved)} & \textbf{"Bored but can't say what I want"} \\
Explicit intent & Search & "Weekend trips near Beijing" \\
\bottomrule
\end{tabularx}
\end{table}

But there is a large middle ground---what we call \textit{vague intent}---where neither tool works. The user senses that the feed has gone stale, feels a restless desire for "something different" maybe lingers on an unfamiliar post for a second longer than usual. The feed will not help: it will keep serving more of the same. The search box will not help either: the user cannot type a query for an interest they have not yet discovered. They are stuck---aware of a problem, unable to act~\cite{brown2020language, devlin2019bert, luo2024integrating, li2023gpt4rec, yang2023palr, liao2024llara, ji2024genrec, li2025survey, deng2025onerec, wei2022chain, kojima2022large}.

This is the gap we target. Our premise is straightforward: if the user cannot tell the system what they want, the system should start the conversation. Rather than waiting for a query that will never come, the AI can look at what the user has been browsing, say what it sees, and offer low-effort options for exploring outward.

\system{} implements this idea. The core inversion is simple: instead of the user telling the AI what they want (the paradigm of every chatbot and search engine), the AI tells the user what it observes and offers to help. The interaction has four stages:

\begin{enumerate}[nosep, leftmargin=*]
    \item \textbf{AI-initiated insight.} The AI analyzes the current feed composition and proactively surfaces an observation (e.g., "Your feed is 70\% food and fashion content. I notice you occasionally linger on travel posts").
    \item \textbf{Low-cost response options.} Instead of requiring free-text input, the AI presents clickable options that capture common vague-intent states (e.g., "Show me more travel," "I'm bored with food content" "Surprise me with something new").
    \item \textbf{Guided narrowing.} If the user selects an option, the AI follows up with one clarifying round---still option-based---to refine the direction without requiring precise articulation.
    \item \textbf{Gradual feed blending.} New content is progressively mixed into the existing feed rather than replacing it wholesale, allowing the user to discover new interests organically.
\end{enumerate}

This paper makes three contributions.
\begin{itemize}[nosep, leftmargin=*]
    \item We identify \textit{vague intent} as an under-supported state in content exploration: users want their feed to change, but cannot yet formulate a specific request. A formative study (\n{}=12) characterizes this experience and informs five design requirements.
    \item We present \system{}, an exploration interface that combines feed-level behavioral reflection, option-based interaction, and gradual feed blending to support low-effort redirection of recommendation feeds.
    \item Through a mixed-design lab study (\n{}=28), we show that this AI-initiated, structured interaction flow supports broader exploration and lower interaction effort than passive feeds, search, and a user-initiated chat interface, while maintaining users' sense of control.
\end{itemize}

\section{Related Work}

\subsection{Filter Bubbles and Content Diversity}

Pariser's \textit{filter bubble} thesis~\cite{pariser2011filter}---that algorithmic personalization narrows what people see---has held up well empirically. Nguyen et al.~\cite{nguyen2014exploring} showed that collaborative filtering reduces content diversity over time, and Bakshy et al.~\cite{bakshy2015exposure} found that on Facebook, algorithmic curation and user self-selection both contribute to ideological homogeneity.

The recommender systems community has attacked the problem algorithmically: diversification re-ranking~\cite{ziegler2005improving, kunaver2017diversity}, serendipity objectives~\cite{ge2010beyond, kotkov2019serendipity}, and multi-objective bandit approaches~\cite{mehrotra2020bandit}. These methods improve aggregate diversity metrics, but they are invisible to the user---the diversification happens behind the curtain, and the user has no say in it. On the HCI side, Lu et al.~\cite{lu2024see} built a multi-agent LLM system to expose users to diverse viewpoints, and Resnick et al.~\cite{resnick2013bursting} outlined strategies for promoting exposure diversity. Both focus on broadening perspectives on topics the user already engages with. Our goal is different: helping users \textit{find entirely new topics} they had not considered at all.

\subsection{User Control in Recommendation Interfaces}

There is solid evidence that giving users control over recommendations improves satisfaction, even at the cost of accuracy~\cite{harper2015putting}, and that inspectable user models increase trust~\cite{knijnenburg2012inspectability}. Tsai and Brusilovsky~\cite{tsai2021effects} added nuance: controllability and explainability interact in sometimes counterintuitive ways in social recommender systems. Jin et al.~\cite{jin2022ucrs} went further with UCRS, a framework offering four granularity levels of user control specifically aimed at filter bubbles. Jhaver et al.~\cite{jhaver2023personalizing} found that even modest control can be acceptable if well-designed.

The shared assumption across this work is that the user knows what adjustment to make. UCRS asks users to specify which topics to increase or decrease; content moderation tools require explicit filter rules. We are interested in the case where users \textit{cannot yet specify what they want}---they only know something feels off. Control mechanisms are useless if you do not know which knob to turn.

\subsection{Conversational Recommendation}

Conversational recommender systems (CRS) elicit preferences through dialogue~\cite{jannach2021survey, christakopoulou2016towards, sun2018conversational}. Hernandez-Bocanegra and Ziegler~\cite{hernandez2023explaining} found that interactive options for examining recommendation explanations outperformed free-form chatbot interfaces. TKGPT~\cite{tkgpt2025} let users "chat with" a recommendation algorithm, which improved algorithmic understanding and sense of agency.

All of these systems share one assumption: the user speaks first. The conversation begins with a stated need, and the system responds with clarifying questions and recommendations. This works fine when the user has at least a rough idea of what they want. It breaks down in the vague-intent scenario.

The closest prior work to ours is InterQuest~\cite{mei2025interquest}, a mixed-initiative framework for conversational search (UIST 2025). InterQuest dynamically models user interests during search, combining system-inferred and user-expressed interests. It is an important step toward mixed-initiative information access---but it operates within the search paradigm. Users have already entered a search context; they have at least a nascent query. We start further upstream (Table~\ref{tab:spectrum}): users who are passively scrolling a feed and have formed no query at all. The AI must not only model interests but \textit{initiate the entire exploration process from scratch}.

\subsection{AI-Initiated and Proactive Interaction}

Mixed-initiative interaction---where both user and system can take the lead---is an old idea in HCI~\cite{horvitz1999principles, allen1999mixed}, but LLMs have made it newly practical. Amershi et al.'s~\cite{amershi2019guidelines} guidelines for human-AI interaction flag two points especially relevant for proactive AI: \textit{G2: Make clear what the system can do} and \textit{G5: Match relevant social norms}. Speaking before being spoken to is a social violation if done badly.

Recent work suggests it can be done well. Yoon et al.~\cite{yoon2020proactive} studied proactive AI expectations at scale (\n{}=272) and found that users in recommendation scenarios actually \textit{preferred} higher proactivity. Lee et al.'s Sensible Agent~\cite{lee2025sensible} (UIST 2025) demonstrated unobtrusive proactive interaction in AR that significantly reduced perceived effort. Drosos et al.~\cite{drosos2025provocations} showed that AI-initiated "provocations" in knowledge work restored critical thinking that passive AI assistance had dampened. (Their between-subjects design, chosen to avoid priming effects in conversational AI studies, informed our own methodology.)

We extend this line to content exploration under vague intent. Unlike Sensible Agent, which reacts to detected activities in AR, \system{} initiates a structured multi-turn dialogue to help users articulate and act on interests they cannot yet name.

\section{Formative Study}

Before building \system{}, we needed to understand three things: whether ordinary users actually perceive filter-bubble staleness (industry intuition says yes, but we wanted evidence), what they currently try to do about it, and whether they would accept an AI that proactively comments on their browsing patterns.

\subsection{Method}

\subsubsection{Participants.}
We recruited 12 participants (7 female, 5 male; ages 21--32, M=25.3) through university mailing lists and social media. All were daily users of content recommendation platforms (Xiaohongshu, Douyin) with self-reported usage of at least 30 minutes per day. No participants had backgrounds in HCI or recommender systems. 

\subsubsection{Procedure.}
Each session lasted approximately 60 minutes and consisted of three parts:

\textbf{Part A: Semi-structured interview (15 min).} We explored participants' everyday experience with recommendation feeds, focusing on: (1) how they characterize their typical browsing experience, (2) whether and how often they perceive content repetitiveness, (3) what strategies they employ when dissatisfied, and (4) whether they have experienced wanting different content but being unable to articulate what.

\textbf{Part B: Prototype think-aloud (30 min).} Participants used our \system{} prototype. They first browsed a feed intentionally biased toward 2--3 content categories (approximately 80\% concentration) for 5 minutes, then interacted with the AI-initiated dialogue feature. We employed a think-aloud protocol, asking participants to verbalize their reactions at each step: when the AI surfaced its analysis, when options appeared, and when new content was blended into the feed.

\textbf{Part C: AI proactivity calibration (15 min).} We presented three scenario cards depicting different AI proactivity levels:
\begin{itemize}[nosep, leftmargin=*]
    \item \textbf{Level 1 (Reactive):} AI analyzes the feed only when the user explicitly requests it via a button.
    \item \textbf{Level 2 (Moderate):} AI automatically surfaces insights after the user has browsed for several minutes (our default design).
    \item \textbf{Level 3 (Eager):} AI comments on the feed composition after every refresh or extended scroll.
\end{itemize}
Participants ranked these levels by preference and discussed the reasoning behind their rankings, including when proactive AI intervention would feel helpful versus intrusive.

\subsubsection{Analysis.}
All sessions were audio-recorded and transcribed. We conducted reflexive thematic analysis following Braun and Clarke~\cite{braun2006thematic}. Two researchers independently coded all transcripts, then collaboratively refined themes through iterative discussion until reaching consensus.

\subsection{Findings}

\subsubsection{Everyone recognizes the problem; nobody has a good fix.}
All 12 participants reported experiencing content repetitiveness---from "almost every session" (P3, P7, P11) to "a few times a week" (P1, P5, P9). The language was vivid: \textit{"It's like the algorithm only knows one side of me"} (P4); \textit{"I'm stuck in a loop---food, skincare, food, skincare, repeat"} (P7).

Their coping strategies were strikingly limited:
\begin{itemize}[nosep, leftmargin=*]
    \item \textbf{Keep scrolling} (8/12): Hope the algorithm eventually surfaces something different. \textit{"I just keep swiping and hope"} (P2).
    \item \textbf{Switch apps} (6/12): Close one app, open another. \textit{"When Douyin gets boring I switch to Xiaohongshu, but honestly it's the same problem there"} (P10).
    \item \textbf{Try to search} (4/12): Open the search bar, then give up. \textit{"I'll open search and just stare at it---I literally don't know what to type"} (P8).
    \item \textbf{Mark "not interested"} (3/12): Use the platform's negative feedback. \textit{"I mark things not interested, but it just replaces them with more of the same category"} (P6).
\end{itemize}

\subsubsection{The core obstacle is articulation, not motivation.}
Nine of 12 participants described what we call the \textit{articulation barrier}: wanting something different but being unable to say what. Two quotes capture it well:

\textit{"It's not that I don't want to explore. I don't even know what I don't know. I can't search for a hobby I haven't heard of"} (P3).

\textit{"The search box assumes I already have an answer. But my question isn't `show me X'---it's `what else is out there that I might like?'{"}} (P11).

This is the finding that shaped everything else in the paper. If the obstacle is articulation rather than motivation, the intervention should not demand articulation. The AI should offer concrete starting points that the user can accept, reject, or tweak.

\subsubsection{The "mirror" effect: AI insights trigger self-awareness.}
During the think-aloud (Part B), participants reacted strongly when the AI surfaced its feed analysis (e.g., "Your feed is 72\% food and fashion. You seem to occasionally pause on travel content"):

\textit{"Oh wow, I didn't realize it was that skewed. That's kind of eye-opening"} (P1).

\textit{"Wait, it noticed I stopped on travel posts? I didn't even notice that myself"} (P9).

Several participants described the AI as reflecting their browsing behavior back to them, making implicit patterns feel more visible and actionable. The AI reflected their behavior back to them in a way that made implicit patterns explicit. Once the pattern was visible, deciding what to change felt much easier.

\subsubsection{Clicking beats typing when you do not know what to type.}
All 12 participants clicked the options rather than typing during their first interaction round:

\textit{"This is exactly what I needed---I don't have to think of the right words, I just click the one that feels right"} (P5).

Four participants wanted the ability to type freely in later rounds, once the AI had helped them narrow a direction: \textit{"Once I know what area I'm interested in, I might want to say something more specific"} (P12). This suggested a design that starts with options and opens up free-text as the conversation progresses.

\subsubsection{Moderate proactivity hits the sweet spot.}
Nine of 12 participants ranked Level 2 (Moderate) first. Level 1 (Reactive) was seen as pointless: \textit{"If I have to press a button, I might as well just search"} (P4). Level 3 (Eager) felt intrusive: \textit{"That would feel like someone looking over my shoulder"} (P7).

The preferred level was context-dependent: during commutes or idle browsing, participants wanted more proactivity; during focused research (e.g., comparing products before a purchase), they wanted less.

\subsection{Design Requirements}

These findings yielded five design requirements:

\begin{description}[nosep, leftmargin=*, font=\normalfont\bfseries]
    \item[DR1: AI-initiated dialogue.] The system should proactively initiate the exploration conversation, rather than waiting for user input, because users in the vague-intent state cannot formulate what to ask (Section~3.2.2).
    \item[DR2: Behavioral insight as conversation opener.] The AI's opening should surface a concrete, data-grounded observation about the user's browsing patterns, serving as a "mirror" that makes implicit habits explicit (Section~3.2.3).
    \item[DR3: Option-based interaction with free-text fallback.] The primary interaction modality should be clickable options that lower the articulation barrier, with free-text input available for users who develop more specific preferences (Section~3.2.4).
    \item[DR4: Gradual content integration.] New content should be blended into the existing feed progressively rather than replacing it wholesale, to reduce jarring transitions and allow organic discovery (supported by participant feedback during think-aloud, where sudden changes felt "disorienting" per P6).
    \item[DR5: Moderate proactivity with contextual sensitivity.] The AI should intervene after a period of browsing (not immediately), and ideally adapt its proactivity level to the user's engagement state (Section~3.2.5).
\end{description}

\section{System Design}

\system{} is a desktop application (with the long-term goal of deployment on general recommendation apps) that displays a content recommendation feed while providing an AI-initiated exploration assistant. We describe the design principles, interaction flow, and implementation.

\subsection{Design Principles}

Three principles guide the design, each grounded in the formative study:

\textbf{P1: The AI speaks first (DR1, DR2).} In a conventional chatbot, the user initiates. \system{} inverts this: the AI opens with a substantive observation about the user's browsing, and the user responds by selecting, rejecting, or refining. The user retains full control over whether to engage---the AI proposes, the user disposes.

\textbf{P2: One click should be enough (DR3).} Each interaction should extract maximum signal from minimum effort. Options represent distinct exploration trajectories, not just topic labels. A single click communicates both a preference and a direction.

\textbf{P3: Exploration is part of browsing, not a separate mode (DR4, DR5).} Content changes are gradual. The AI's presence is a floating affordance that does not obstruct the feed. The conversation panel slides in alongside the feed, so users can continue scrolling while chatting.

\subsection{Interaction Flow}

\begin{figure}[t]
\centering

\begin{subfigure}{0.48\columnwidth}
    \includegraphics[width=\linewidth]{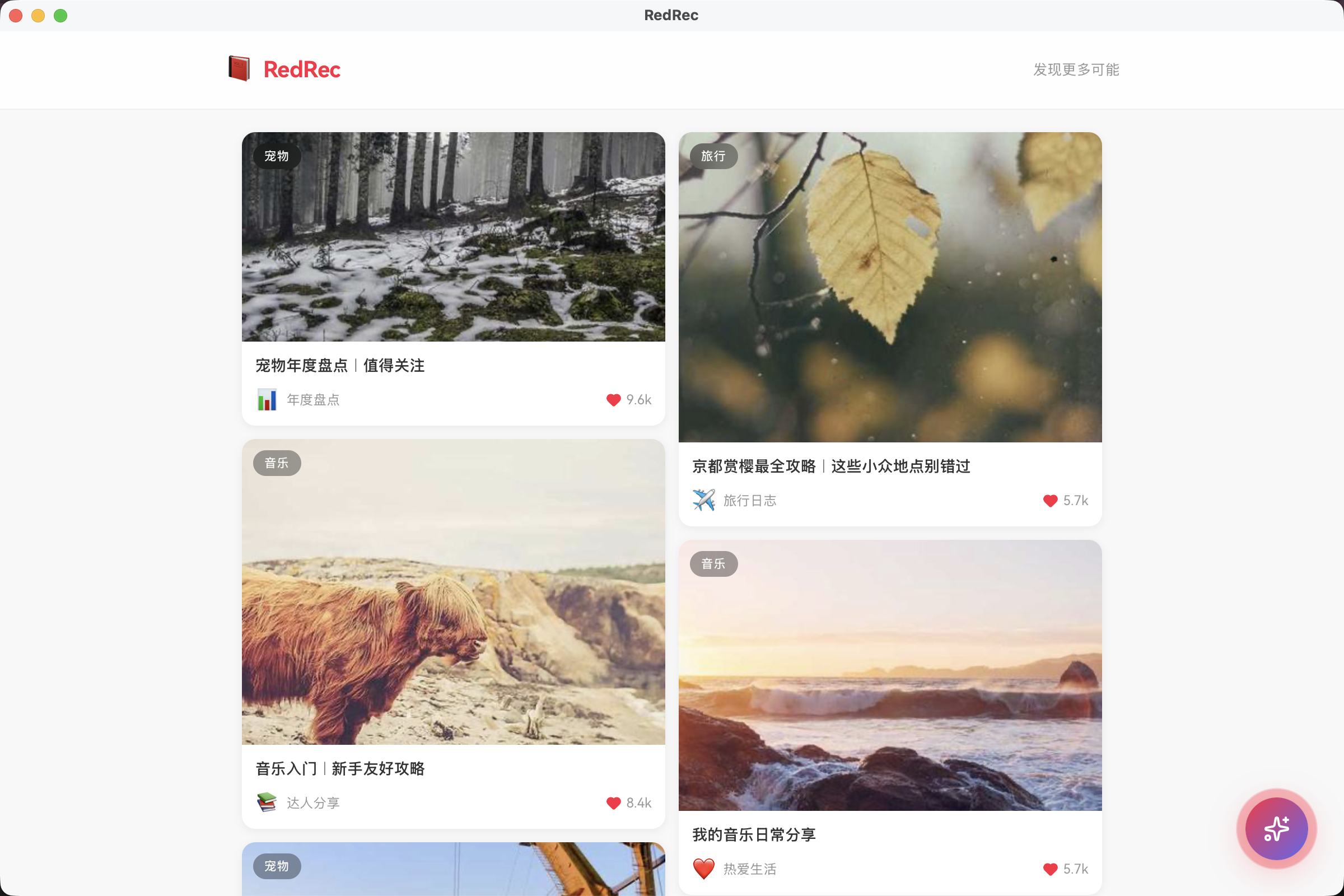}
    \caption{AI-initiated insight surfaces the feed composition}
\end{subfigure}
\hfill
\begin{subfigure}{0.48\columnwidth}
    \includegraphics[width=\linewidth]{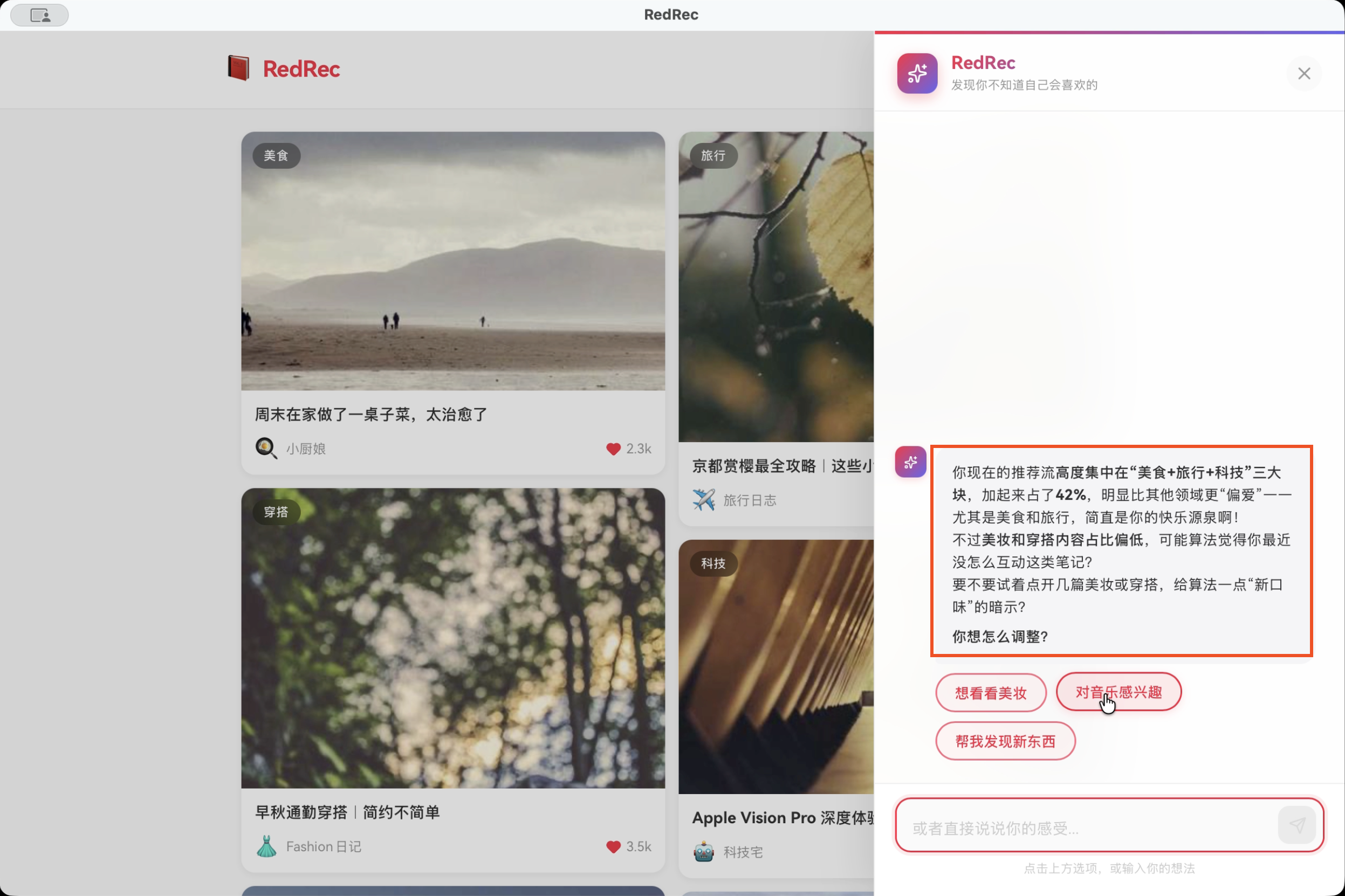}
    \caption{clickable options offer low-cost exploration directions}
\end{subfigure}

\vspace{1ex}

\begin{subfigure}{0.48\columnwidth}
    \includegraphics[width=\linewidth]{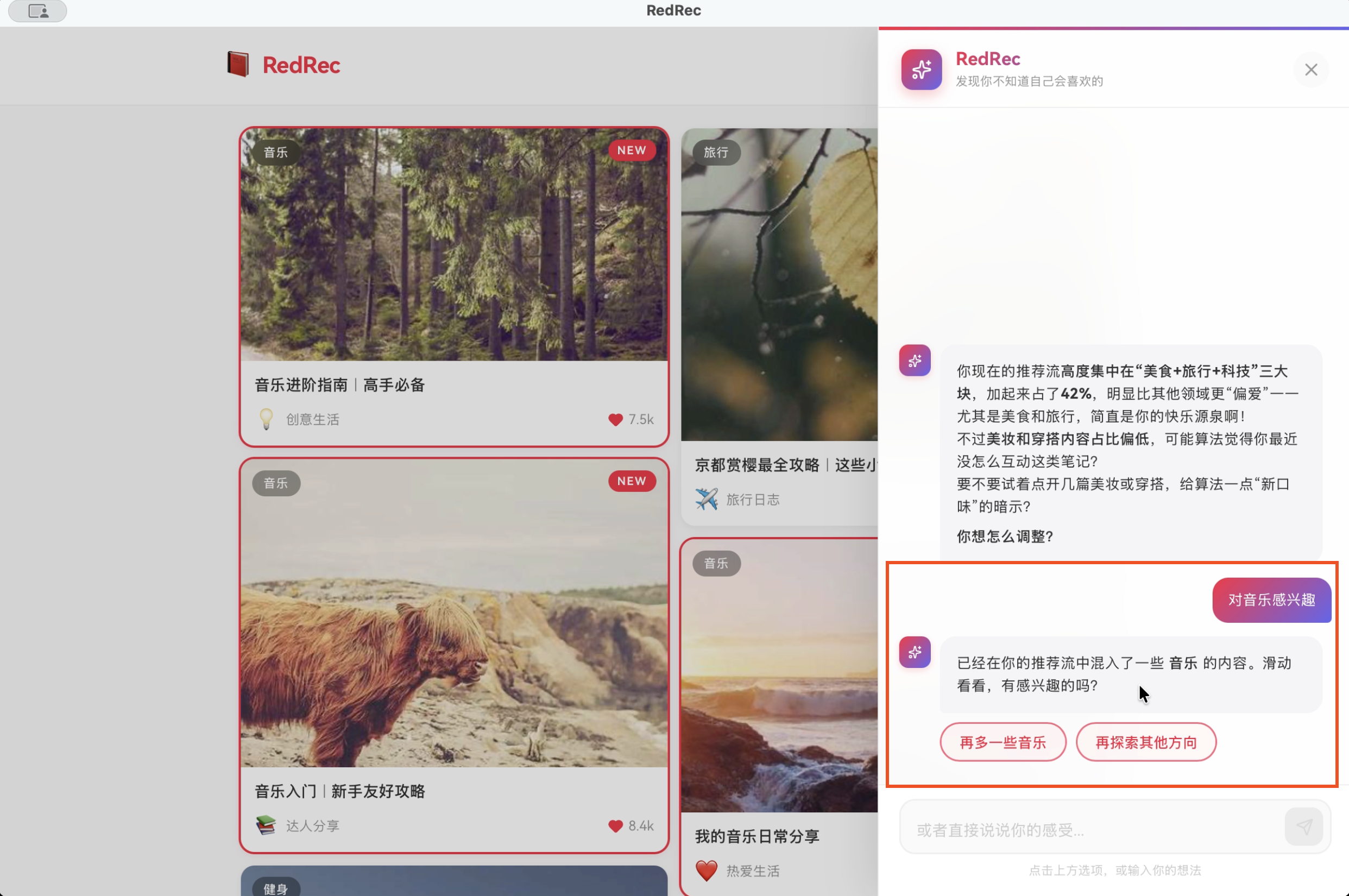}
    \caption{one follow-up question narrows the direction}
\end{subfigure}
\hfill
\begin{subfigure}{0.48\columnwidth}
    \includegraphics[width=\linewidth]{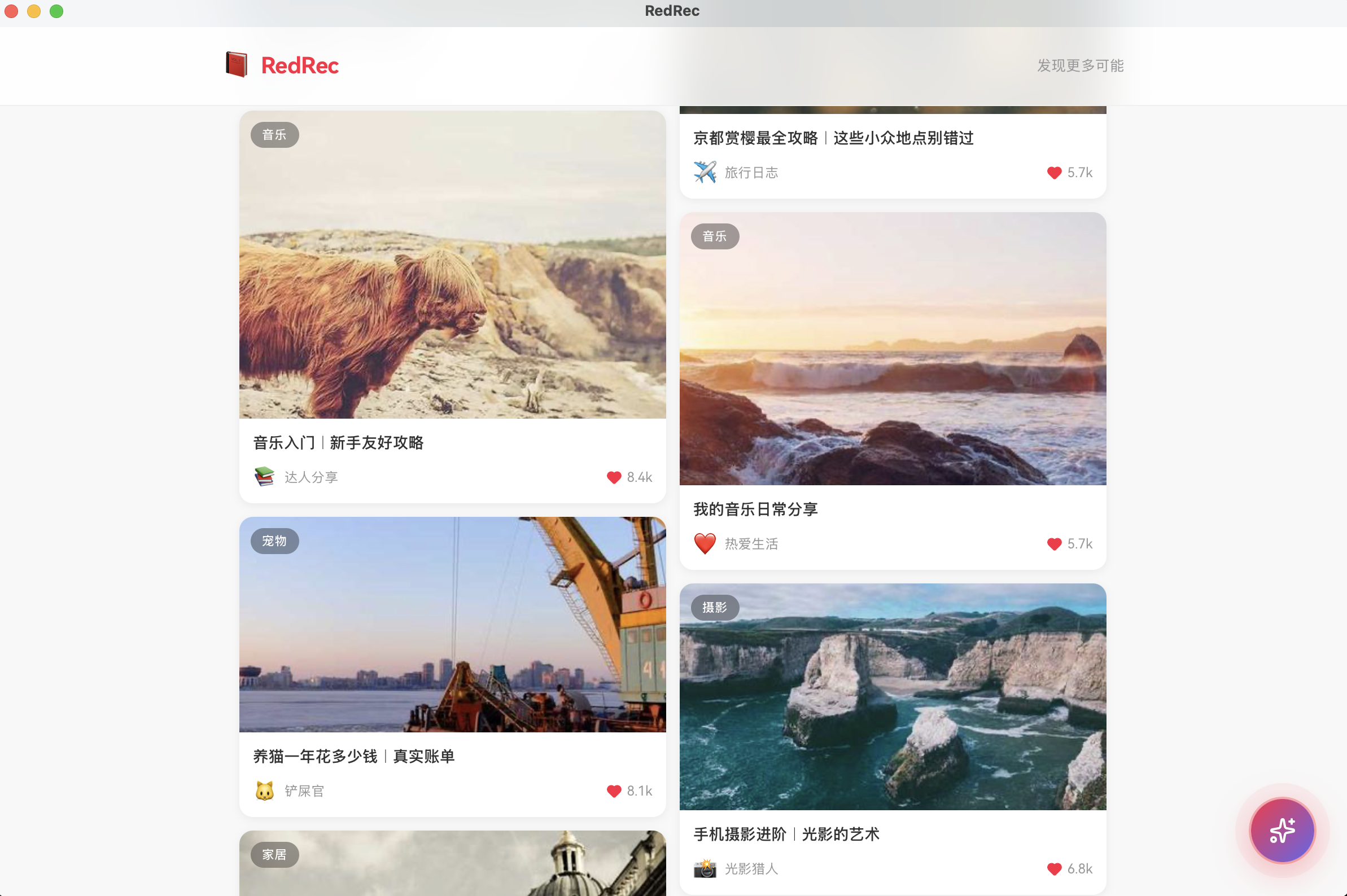}
    \caption{new content blends gradually into the feed}
\end{subfigure}

\caption{The four-stage interaction flow of \system{}.}
\label{fig:interaction-flow}
\end{figure}

The interaction has four stages (Figure~\ref{fig:interaction-flow}).

\subsubsection{Stage 1: AI-Initiated Insight.}
After the user has scrolled through roughly 20 items (configurable), a floating button appears with a subtle animation. Tapping it triggers a feed analysis: the AI computes the category distribution, identifies dominant and underrepresented categories, and picks up behavioral signals like dwell time on specific content types. It then opens the conversation panel with a concrete observation:

\begin{quote}
\textit{"Your feed is about 75\% food and fashion. I noticed you tend to pause on travel-related posts. Want to explore that direction, or something completely different?"}
\end{quote}

This is the "mirror" from the formative study (\textbf{DR2})---grounded in data, not speculation.

\subsubsection{Stage 2: Low-Cost Response Options.}
Below the insight, the AI presents 3--4 clickable options (\textbf{DR3}), each representing a distinct exploration direction. For example:

\begin{itemize}[nosep, leftmargin=*]
    \item \textit{"Yes, show me more travel content"} (pursue the detected signal)
    \item \textit{"Less food---it's getting repetitive"} (reduce a dominant category)
    \item \textit{"Surprise me with something completely new"} (maximum exploration)
    \item A text input field for free-form responses
\end{itemize}

Options are generated by the LLM based on the feed analysis, so they are contextually specific rather than generic. Zero typing required---but the text field is there for users who want it.

\subsubsection{Stage 3: Guided Narrowing.}
If the user selects an option, the AI follows up with at most one clarifying question---still option-based:

\begin{quote}
\textit{"For travel, are you more into:}
\begin{itemize}[nosep, leftmargin=*]
    \item \textit{Weekend getaways nearby}
    \item \textit{Long-trip planning and itineraries}
    \item \textit{Vicarious travel---beautiful photos and stories"}
\end{itemize}
\end{quote}

This moves the user from a vague direction to a specific one without ever requiring them to formulate a query from scratch. The user's role throughout is \textit{editor}, not \textit{author}.

\subsubsection{Stage 4: Gradual Feed Blending.}
Once a direction is established, \system{} blends new content into the existing feed \textit{gradually} (\textbf{DR4}). Rather than replacing everything, it substitutes roughly 20--30\% of items per refresh cycle. The user continues scrolling and encounters new content interspersed with familiar types---no jarring page reload, no "you are now in discovery mode" modal.

The AI confirms with a brief message: \textit{"I've started mixing in travel content. Keep scrolling---let me know if you want to adjust."}

\subsection{Implementation}

\system{} is a Tauri 2 desktop application: React 19 + TypeScript frontend, Rust backend. State management uses Zustand; UI components come from Ant Design. The Rust backend handles LLM calls via \texttt{reqwest}/\texttt{tokio}.

The LLM is Qwen3-VL-235B, accessed through an API. Prompts are designed for structured output: the insight prompt returns a natural-language analysis with category distributions; the option prompt returns 3--4 contextually relevant choices; the content prompt returns structured note objects matching the feed schema.

The recommendation algorithm for information flow relies on an end-to-end generative recommendation system implemented via SAGE~\cite{xie2026sage}. This system is more receptive to conversational text as input and can adjust recommendation results in a timely manner. Rather than simply delivering more content that users need, our goal is to provide more content that users will genuinely like. If a user clicks on newly recommended content, the system will grant a larger reward.

For the experiment, a condition-switching module toggles between the four interface variants (FEED, SEARCH, USER-CHAT, AI-INIT) via a config flag. Comprehensive behavioral logging records timestamped events for scrolling, clicking, conversation turns, option selections, and content exposure durations.

\section{Evaluation}

We ran a controlled lab study comparing \system{} against three baselines. Three research questions:

\begin{description}[nosep, leftmargin=*, font=\normalfont\bfseries]
    \item[RQ1:] Does AI-initiated interaction help users discover content beyond their filter bubble, compared to passive feeds and search?
    \item[RQ2:] How does an AI-initiated, structured dialogue flow compare with a user-initiated chat interface for content exploration?
    \item[RQ3:] How do users perceive AI-initiated intervention in terms of interaction effort, control, and agency?
\end{description}

\subsection{Study Design}

We used a \textbf{mixed design} with one between-subjects and one within-subjects factor:

\begin{itemize}[nosep, leftmargin=*]
    \item \textbf{Between-subjects factor: Dialogue mode.} Participants were randomly assigned to one of two groups:
    \begin{itemize}[nosep, leftmargin=*]
        \item \textbf{AI-INIT group}: experienced the full \system{} with AI-initiated interaction.
        \item \textbf{USER-CHAT group}: experienced a variant where the chat panel was available but the AI did not initiate---it displayed "How can I help you?" and waited for user input. The AI's response capability (same LLM, same knowledge) was identical; only the initiation dynamic differed.
    \end{itemize}
    \item \textbf{Within-subjects factor: Baseline conditions.} All participants experienced both baseline conditions (FEED and SEARCH) in counterbalanced order before their assigned dialogue condition.
\end{itemize}

Why mixed? The AI-INIT vs.\ USER-CHAT comparison must be between-subjects because AI-initiated dialogue creates an irreversible priming effect: once the AI has pointed out your filter bubble, you cannot unsee it, which would contaminate a subsequent user-initiated condition. Drosos et al.~\cite{drosos2025provocations} made the same choice for the same reason. The baseline conditions (FEED, SEARCH) involve no AI dialogue and carry no priming risk, so we compare them within-subjects for higher statistical power.

\subsection{Conditions}

\begin{table}[t]
\centering
\caption{Summary of the four experimental conditions.}
\label{tab:conditions}
\small
\begin{tabular}{p{1.8cm}p{5.8cm}}
\toprule
\textbf{Condition} & \textbf{Description} \\
\midrule
FEED & Standard two-column masonry feed with pull-to-refresh. No exploration tools. \\
\addlinespace
SEARCH & Feed + a search bar supporting keyword and category-based filtering. Search results supplement or replace the current feed. \\
\addlinespace
USER-CHAT & Feed + a chat panel. The AI does not initiate; it displays a prompt ("How can I help you?") and responds only after user input. No options are auto-generated. Same LLM backend as AI-INIT. \\
\addlinespace
AI-INIT & Full \system{}: AI proactively analyzes the feed, surfaces insights, offers clickable options, follows up with guided narrowing, and gradually blends new content. \\
\bottomrule
\end{tabular}
\end{table}

Table~\ref{tab:conditions} summarizes the four conditions. USER-CHAT is the minimal contrast to AI-INIT: identical UI, identical LLM, but the user must initiate. This isolates the contribution of \textit{AI-initiated structure}---the insight, the options, the guided narrowing---from the mere \textit{availability} of a chat-capable AI.

\subsection{Participants}

We recruited 28 participants (16F, 12M; ages 19--34, $M$=24.8, $SD$=3.6) via university mailing lists and campus boards. All used at least one recommendation platform (Xiaohongshu or Douyin) for $\geq$30 min/day. We excluded anyone with HCI, recommender systems, or AI research backgrounds. None had participated in Study 1.

Random assignment to AI-INIT ($n$=14) and USER-CHAT ($n$=14), balanced on gender, age, and daily app usage. Compensation: gift card equivalent to \$10 USD.

\subsection{Materials}

\subsubsection{Content corpus.}
We curated 320 real Xiaohongshu posts spanning 14 categories (food, fashion, skincare, travel, fitness, home decor, photography, technology, reading, pets, outdoor activities, art, music, parenting). Each item had a title, cover image, author name, engagement count, and category label, manually verified for quality.

\subsubsection{Biased initial feeds.}
Three biased feeds, each 35 items at $\sim$80\% concentration in 2--3 categories:

\begin{itemize}[nosep, leftmargin=*]
    \item Feed A: 80\% food + fashion, 20\% scattered
    \item Feed B: 80\% skincare + fitness, 20\% scattered
    \item Feed C: 80\% home decor + photography, 20\% scattered
\end{itemize}

Feed assignments were counterbalanced across conditions and groups to ensure that no participant encountered the same feed content twice and that each feed configuration appeared equally often in each condition.

\subsection{Procedure}

Each experimental session lasted approximately 90 minutes and followed this protocol:

\begin{enumerate}[nosep, leftmargin=*]
    \item \textbf{Pre-experiment (10 min):} Informed consent, demographics questionnaire, and a baseline "filter bubble awareness" survey.
    
    \item \textbf{Condition 1---Baseline (25 min):} One of FEED or SEARCH (counterbalanced within group).
    \begin{itemize}[nosep, leftmargin=*]
        \item \textit{Warm-up phase (5 min):} Browse the biased feed without any task instruction.
        \item \textit{Free exploration (15 min):} Instruction: "If you have any thoughts about the current content---if it feels repetitive, if you want something different, or if you have any vague sense that you'd like to change things---feel free to use whatever tools the interface provides. There is no specific goal."
        \item \textit{Condition questionnaire (5 min):} Fill out measures for this condition.
    \end{itemize}
    
    \item \textbf{Break (5 min).}
    
    \item \textbf{Condition 2---Baseline (25 min):} The other baseline condition (FEED or SEARCH), with a different biased feed. Same structure as Condition 1.
    
    \item \textbf{Break (5 min).}
    
    \item \textbf{Condition 3---Experimental (25 min):} AI-INIT or USER-CHAT (between-subjects assignment), with the third biased feed. Same structure as Condition 1.
    
    \item \textbf{Post-experiment interview (15--20 min):} Semi-structured interview comparing experiences across all three conditions (see Section~\ref{sec:interview}).
\end{enumerate}

All sessions took place in a quiet lab room with screen recording. The system logged all interactions with timestamps.

\subsection{Measures}

\subsubsection{Primary measures---Exploration effectiveness (RQ1).}
\begin{itemize}[nosep, leftmargin=*]
    \item \textbf{Exploration breadth:} Number of distinct content categories the participant browsed during the 15-minute exploration phase (system log).
    \item \textbf{Diversity gain ($\Delta H$):} Change in Shannon entropy~\cite{shannon1948mathematical} of the category distribution between the end of the warm-up phase and the end of the exploration phase. $\Delta H = H_{\text{post}} - H_{\text{pre}}$, where $H = -\sum_i p_i \log_2 p_i$ over category proportions in the viewed content (system log).
    \item \textbf{Bubble-breaking rate:} Proportion of content categories that were underrepresented ($<$5\% of initial feed) but were browsed during exploration (system log).
    \item \textbf{Serendipity:} 4-item 7-point Likert scale adapted from Kotkov et al.~\cite{kotkov2019serendipity}: (1) "I discovered content I didn't expect but found interesting"; (2) "The system helped me encounter content I would not have searched for"; (3) "I was pleasantly surprised by some of the new content"; (4) "This experience broadened my content horizons."
\end{itemize}

\subsubsection{Secondary measures---Interaction efficiency (RQ2).}
\begin{itemize}[nosep, leftmargin=*]
    \item \textbf{Perceived interaction effort:} 3-item 7-point Likert scale adapted from Sensible Agent~\cite{lee2025sensible}: (1) "Using this tool required a lot of mental effort" (reverse); (2) "It was easy to express what I wanted"; (3) "Getting to content I liked required too many steps" (reverse).
    \item \textbf{Expression cost:} Total characters typed by the participant during the exploration phase (system log; option clicks count as 0 characters).
    \item \textbf{Time to first discovery:} Seconds from the start of the exploration phase to the first instance of viewing content from a previously underrepresented category ($<$5\% of initial feed) for at least 2 seconds (system log).
    \item \textbf{Tool engagement rate:} Whether the participant actively used the available tool (search/chat) within the first 5 minutes of the exploration phase (system log).
\end{itemize}

\subsubsection{Experiential measures (RQ3).}
\begin{itemize}[nosep, leftmargin=*]
    \item \textbf{Perceived control:} 4-item 7-point Likert scale: (1) "I felt in control of the browsing direction"; (2) "Content changes matched my expectations"; (3) "I could change direction at any time"; (4) "The final content mix reflected my intentions."
    \item \textbf{Sense of agency:} 3-item 7-point Likert scale adapted from TKGPT~\cite{tkgpt2025}: (1) "I understood why the system recommended this content"; (2) "I felt like I was actively exploring rather than passively receiving"; (3) "I had influence over the recommendations."
    \item \textbf{System usability:} System Usability Scale (SUS)~\cite{brooke1996sus}, 10 items.
    \item \textbf{AI proactivity acceptability:} 4-item 7-point Likert scale (AI-INIT and USER-CHAT groups only): (1) "The AI intervened at an appropriate time"; (2) "The AI's suggestions were helpful"; (3) "The AI felt intrusive" (reverse); (4) "I would have preferred the AI to be more proactive."
\end{itemize}

\subsubsection{Behavioral log measures.}
The system automatically recorded: content exposure events (item ID, timestamp, duration visible in viewport), scroll depth, click events, conversation logs (each turn with timestamp, content, and whether it was a click on an option or free-text input), and feed composition changes over time.

\subsection{Post-Experiment Interview}
\label{sec:interview}

Semi-structured interviews with all 28 participants covered:

\textbf{All participants:}
\begin{enumerate}[nosep, leftmargin=*]
    \item Describe each of the three interfaces you experienced in one sentence.
    \item Which interface best helped you discover content you didn't previously know you'd enjoy? Why?
    \item Was there a moment where you thought "I didn't know I'd be interested in that"? In which condition?
    \item If this feature existed in your daily recommendation app, when would you want it to appear? How often?
\end{enumerate}

\textbf{AI-INIT group only:}
\begin{enumerate}[nosep, leftmargin=*, start=5]
    \item What was your first reaction when the AI proactively described your browsing patterns?
    \item Among the AI's suggested options, were there any that captured exactly what you were feeling? Any that felt off?
    \item The content changed gradually rather than all at once---did you notice? How did that feel?
\end{enumerate}

\textbf{USER-CHAT group only:}
\begin{enumerate}[nosep, leftmargin=*, start=5]
    \item When you opened the chat panel, what was your first instinct to say?
    \item Were there moments when you wanted to say something but didn't know how to express it?
    \item Would you have preferred the AI to speak first rather than waiting for you?
\end{enumerate}

\subsection{Analysis Plan}

\subsubsection{Quantitative analysis.}
For within-subjects comparisons (FEED vs.\ SEARCH, within each group), we use paired $t$-tests (or Wilcoxon signed-rank tests for non-normal Likert data). For the critical between-subjects comparison (AI-INIT vs.\ USER-CHAT), we use independent-samples $t$-tests (or Mann-Whitney $U$ tests). For omnibus comparisons across all conditions, we use mixed ANOVA with dialogue mode (AI-INIT, USER-CHAT) as the between-subjects factor and baseline condition (FEED, SEARCH) as the within-subjects factor, with the experimental condition incorporated through planned contrasts. Effect sizes are reported as Cohen's $d$ for pairwise comparisons and partial $\eta^2$ for ANOVA. Multiple comparisons are corrected using Bonferroni adjustment.

\subsubsection{Qualitative analysis.}
Interview transcripts are analyzed using reflexive thematic analysis~\cite{braun2006thematic}. Two researchers independently generate initial codes from the data, then collaboratively develop, review, and refine themes through iterative discussion. We report inter-coder reliability using Cohen's $\kappa$ on a subset of transcripts, targeting $\kappa \geq 0.7$.

\subsection{Results}
\label{sec:results}

We report quantitative results first, then behavioral log analysis, then qualitative findings. Unless otherwise noted, we use independent-samples $t$-tests for between-subjects comparisons (AI-INIT vs.\ USER-CHAT) and paired $t$-tests for within-subjects comparisons (FEED vs.\ SEARCH). Where Likert data violated normality (Shapiro-Wilk $p<.05$), we confirmed results with non-parametric equivalents; conclusions were identical in all cases, so we report parametric statistics for consistency. All pairwise $p$-values are Bonferroni-corrected.

\subsubsection{Exploration effectiveness (RQ1).}

Table~\ref{tab:exploration} summarizes the exploration metrics across all four conditions.

\begin{table}[t]
\centering
\small
\begin{threeparttable}
\caption{Exploration effectiveness across conditions. Values are $M$ ($SD$).}
\label{tab:exploration}
\begin{tabularx}{0.5\textwidth}{>{\raggedright\arraybackslash}Xcccc}
\toprule
\textbf{Measure} & \textbf{FEED} & \textbf{SEARCH} & \textbf{USER} & \textbf{AI} \\
\midrule
Breadth (\# cat.) & 2.8 (0.9) & 4.6 (1.4)$^a$ & 4.1 (1.7)$^a$ & 7.4 (1.8)$^{abc}$ \\
Diversity gain ($\Delta H$) & 0.12 (0.18) & 0.58 (0.31)$^a$ & 0.43 (0.35)$^a$ & 1.24 (0.41)$^{abc}$ \\
Bubble-breaking (\%) & 8.2 (7.1) & 22.6 (12.4)$^a$ & 18.9 (14.3) & 48.7 (15.2)$^{abc}$ \\
Serendipity (1--7) & 2.86 (0.93) & 3.64 (1.02)$^a$ & 4.31 (1.18)$^{ab}$ & 5.72 (0.84)$^{abc}$ \\
\bottomrule
\end{tabularx}
\begin{tablenotes}[flushleft]
\footnotesize
\item Breadth, diversity gain, and bubble-breaking were computed from logs; serendipity was self-reported. Superscripts indicate significant pairwise differences ($p<.05$, Bonferroni-corrected): $^a$ vs.\ FEED, $^b$ vs.\ SEARCH, $^c$ vs.\ USER-CHAT.
\end{tablenotes}
\end{threeparttable}
\end{table}

\textbf{Exploration breadth.} A mixed ANOVA with dialogue mode (AI-INIT, USER-CHAT) as the between-subjects factor and baseline (FEED, SEARCH) as the within-subjects factor revealed a significant main effect of condition, $F(2.1, 54.6)$=42.3, $p<.001$, $\eta^2_p$=.62 (Greenhouse-Geisser corrected). Planned contrasts showed that AI-INIT participants explored significantly more categories ($M$=7.4, $SD$=1.8) than USER-CHAT ($M$=4.1, $SD$=1.7), $t(26)$=4.97, $p<.001$, $d$=1.88; SEARCH ($M$=4.6, $SD$=1.4), $t(13)$=5.21, $p<.001$, $d$=1.87; and FEED ($M$=2.8, $SD$=0.9), $t(13)$=9.42, $p<.001$, $d$=3.28. USER-CHAT did not significantly outperform SEARCH, $t(26)$=0.94, $p$=.36.

\textbf{Diversity gain.} AI-INIT produced the largest increase in Shannon entropy ($M$=1.24 bits, $SD$=0.41), significantly exceeding USER-CHAT ($M$=0.43, $SD$=0.35), $t(26)$=5.55, $p<.001$, $d$=2.12. FEED participants showed near-zero diversity gain ($M$=0.12, $SD$=0.18), confirming that passive scrolling does not break filter bubbles.

\textbf{Serendipity.} AI-INIT scored highest on the serendipity scale ($M$=5.72, $SD$=0.84), significantly above USER-CHAT ($M$=4.31, $SD$=1.18), $t(26)$=3.61, $p$=.001, $d$=1.37. Both chat conditions outperformed the baselines ($p$s$<.05$), and SEARCH outperformed FEED ($p<.05$).

\subsubsection{Interaction efficiency (RQ2).}

Table~\ref{tab:efficiency} summarizes efficiency metrics for the two chat conditions. (FEED and SEARCH lack conversational interaction, so expression cost and conversation metrics do not apply; we report tool engagement and time to first discovery for all conditions.)

\begin{table}[t]
\centering
\small
\begin{threeparttable}
\caption{Interaction efficiency across conditions.}
\label{tab:efficiency}
\begin{tabularx}{0.5\textwidth}{>{\raggedright\arraybackslash}Xcccc}
\toprule
\textbf{Measure} & \textbf{FEED} & \textbf{SEARCH} & \textbf{USER} & \textbf{AI} \\
\midrule
Perceived effort (1--7, $\downarrow$) & --- & --- & 4.52 (1.21) & 2.74 (0.89)$^c$ \\
Expression cost (chars) & 0 & 18.4 (22.1) & 73.2 (41.6) & 6.1 (12.8)$^c$ \\
Time to 1st discovery (s) & 412 (187) & 198 (94)$^a$ & 247 (126)$^a$ & 89 (43)$^{abc}$ \\
Tool engagement (\%, first 5 min) & --- & 71.4 & 50.0 & 100.0 \\
\bottomrule
\end{tabularx}
\begin{tablenotes}[flushleft]
\footnotesize
\item Expression cost and conversation depth are reported for chat conditions only. Time to first discovery and tool engagement are reported for all conditions.
\end{tablenotes}
\end{threeparttable}
\end{table}

\textbf{Perceived interaction effort.} AI-INIT participants reported significantly lower effort ($M$=2.74, $SD$=0.89) than USER-CHAT ($M$=4.52, $SD$=1.21), $t(26)$=-4.40, $p<.001$, $d$=1.67. This is a large effect: the AI-initiated structure roughly halved the perceived effort of using the conversational tool.

\textbf{Expression cost.} AI-INIT participants typed a median of 0 characters (mean 6.1, mostly from the 3 participants who used free-text in later rounds). USER-CHAT participants typed a median of 73 characters ($M$=73.2, $SD$=41.6). This 12$\times$ difference reflects the option-based design: clicking replaces typing.

\textbf{Time to first discovery.} AI-INIT participants encountered underrepresented content nearly 3$\times$ faster ($M$=89s, $SD$=43) than USER-CHAT ($M$=247s, $SD$=126), $t(26)$=-4.48, $p<.001$, $d$=1.68. FEED participants took the longest ($M$=412s, $SD$=187), and four FEED participants never encountered underrepresented content at all during the 15-minute window.

\textbf{Tool engagement.} All 14 AI-INIT participants engaged with the conversation (by design---the AI initiates). In USER-CHAT, only 7 of 14 (50\%) opened the chat panel within the first 5 minutes. Of the 7 who did not, 4 later reported they "didn't know what to say" and 3 "didn't notice the chat icon." This gap---50\% vs.\ 100\% engagement---is arguably the most practically important finding: the best exploration tool is useless if half your users never start using it.

\subsubsection{User experience (RQ3).}

\begin{table}[t]
\centering
\caption{User experience measures for the two chat conditions (1--7 Likert scales). SUS is on the standard 0--100 scale.}
\label{tab:ux}
\small
\begin{tabular}{lcc}
\toprule
\textbf{Measure} & \textbf{USER-CHAT} & \textbf{AI-INIT} \\
\midrule
Perceived control (1--7) & 4.89 (1.08) & 5.14 (0.92) \\
Sense of agency (1--7) & 4.07 (1.22) & 5.43 (0.87)$^*$ \\
SUS (0--100) & 68.2 (12.4) & 78.6 (9.1)$^*$ \\
AI proactivity acceptability (1--7) & 4.18 (1.31) & 5.61 (0.96)$^*$ \\
\bottomrule
\multicolumn{3}{l}{\small $^*p<.05$, independent-samples $t$-test.}
\end{tabular}
\end{table}

\textbf{Perceived control.} No significant difference between AI-INIT ($M$=5.14, $SD$=0.92) and USER-CHAT ($M$=4.89, $SD$=1.08), $t(26)$=0.67, $p$=.51, $d$=0.25. This is an important null result: the AI speaking first did not make users feel less in control.

\textbf{Sense of agency.} AI-INIT participants reported significantly higher agency ($M$=5.43, $SD$=0.87) than USER-CHAT ($M$=4.07, $SD$=1.22), $t(26)$=3.38, $p$=.002, $d$=1.28. This is counterintuitive---the condition where the AI leads produces higher \textit{user} agency---but the qualitative data (Section~5.8.5) clarifies: the AI's framing helped users feel they were "actively exploring" rather than "passively scrolling," even though the AI initiated.

\textbf{SUS.} AI-INIT scored 78.6 ($SD$=9.1), above the "good" threshold of 68~\cite{brooke1996sus}. USER-CHAT scored 68.2 ($SD$=12.4), right at the threshold. The difference was significant, $t(26)$=2.54, $p$=.017, $d$=0.96.

\textbf{AI proactivity acceptability.} AI-INIT participants found the AI's proactivity highly acceptable ($M$=5.61, $SD$=0.96), significantly above USER-CHAT's rating of the AI's passivity ($M$=4.18, $SD$=1.31), $t(26)$=3.28, $p$=.003, $d$=1.24. Notably, 11 of 14 USER-CHAT participants answered "yes" when asked post-experiment whether they would have preferred the AI to speak first.

\subsubsection{Behavioral patterns.}

System logs revealed several patterns that complement the self-report data:

\textbf{Option vs.\ free-text usage.} In AI-INIT, 13 of 14 participants (93\%) used only option clicks in their first conversation round. In the second round (guided narrowing), 11 of 14 (79\%) continued using options. By the third interaction cycle (if they re-engaged), 4 of 9 re-engaging participants switched to free-text. This confirms the formative study finding: options scaffold the initial interaction, and free-text becomes useful once the user has a direction.

\textbf{Dwell time on new content.} Participants in AI-INIT spent significantly longer viewing newly blended content ($M$=4.8s per item, $SD$=2.1) compared to original feed content ($M$=2.3s, $SD$=1.1), paired $t(13)$=4.12, $p$=.001. In FEED, dwell times were uniformly low across all content ($M$=2.1s, $SD$=0.9).

\textbf{Conversation depth.} AI-INIT conversations averaged 3.2 turns ($SD$=1.1) before participants were satisfied with the feed direction. The distribution was right-skewed: 5 participants reached a satisfactory state in 2 turns (insight + one option click), while 2 participants engaged in 5+ turns, iteratively refining their exploration direction.

\textbf{Scroll velocity change.} After feed adjustment in AI-INIT, scroll velocity decreased by 23\% on average (from $M$=142 to $M$=109 px/s), suggesting participants shifted from rapid scanning to more attentive browsing when encountering new content types.

\subsubsection{Qualitative findings.}

Reflexive thematic analysis of the 28 post-experiment interviews (inter-coder $\kappa$=0.81 on a 30\% subsample) produced five themes. We report each with representative quotes.

\textbf{Theme 1: "The mirror effect."} AI-INIT participants consistently described the AI's initial insight as a moment of self-recognition. The AI did not tell them what to do; it told them what they were already doing, which catalyzed their own decision-making:

\textit{"It was like someone held up a mirror. I knew my feed was repetitive, but seeing `75\% food and fashion' made it concrete. After that, I knew exactly what I wanted to change."} (P16, AI-INIT)

\textit{"I never actually counted what I was seeing. When the AI said it, I thought---yeah, that's right, and I don't want that."} (P22, AI-INIT)

\textbf{Theme 2: "I didn't know I'd like that."} The strongest evidence for serendipitous discovery came from participants who followed the AI's detected signals (e.g., dwell time on travel posts) and were surprised by their own interest:

\textit{"It said I paused on outdoor posts. I didn't even notice I was doing that. But when it showed me hiking and camping content, I was like---yes, this is actually what I want."} (P19, AI-INIT)

\textit{"I went from `I'm bored' to `oh, I'm interested in photography now?' in about two minutes."} (P25, AI-INIT)

\textbf{Theme 3: "The blank page problem."} USER-CHAT participants described a distinctive struggle when confronted with an open-ended chat prompt. The chat panel showed "How can I help you?" and waited---and many users did not know how to begin:

\textit{"I stared at the chat box for a while. I wanted to say something, but what? `Make my feed better'? That felt too vague. So I just kept scrolling."} (P15, USER-CHAT)

\textit{"If it had given me some starting points, I would have engaged immediately. The blank input box was intimidating."} (P21, USER-CHAT)

Five of 14 USER-CHAT participants explicitly said they would have preferred the AI to speak first. This mirrors the formative study's articulation barrier finding, now observed under controlled conditions.

\textbf{Theme 4: "Options as scaffolding."} AI-INIT participants valued the clickable options not just for their low effort, but because the options \textit{taught them what was possible}:

\textit{"I wouldn't have thought to say `surprise me with something new'---but when I saw it as an option, I thought, yeah, that's exactly it."} (P18, AI-INIT)

\textit{"The options showed me what kinds of things I could ask for. It's like a menu at a restaurant you've never been to."} (P27, AI-INIT)

\textbf{Theme 5: "Gradual feels natural."} Participants noticed and appreciated the gradual content blending:

\textit{"It didn't feel like the whole page changed. It was more like new things started appearing between the usual stuff. That felt natural, not jarring."} (P20, AI-INIT)

Two participants (P17, P24) said they would have preferred a faster transition, suggesting that the optimal blending rate may be a matter of individual preference.

\section{Discussion}

\subsection{Why "AI Speaks First" Works}

The results are clear on the core question: having the AI initiate outperforms having the AI wait. AI-INIT participants explored nearly twice as many categories as USER-CHAT, discovered underrepresented content 3$\times$ faster, and typed 12$\times$ fewer characters. But the \textit{why} matters more than the \textit{what}, because it suggests when this pattern will generalize.

The key mechanism is the shift from \textit{author} to \textit{editor}. In USER-CHAT, the user must produce a request from nothing---a blank text box and a vague sense of dissatisfaction. In AI-INIT, the user reacts to a concrete observation. "Your feed is 75\% food and fashion" is language the user can agree with, push back on, or refine. Producing language is hard; editing language is easy. This is why AI-INIT cut perceived effort in half, and why 50\% of USER-CHAT participants never even opened the chat.

The author-to-editor shift should generalize to any domain where users experience vague dissatisfaction but cannot articulate a specific request: playlist fatigue in music streaming, reading ruts in digital libraries, routine stagnation in fitness apps. The precondition is that the system has enough behavioral data to say something substantive---a generic "How can I help?" is not an AI-initiated interaction, it is a chatbot default.

\subsection{Proactivity Without Overstepping}

The fear with proactive AI is always the same: it will feel intrusive~\cite{amershi2019guidelines, yoon2020proactive}. Our data suggests this fear is manageable but real. AI-INIT scored 5.61/7 on proactivity acceptability, and---crucially---perceived control did not differ from USER-CHAT ($p$=.51). But two participants in the formative study flagged Level 3 (Eager) proactivity as "someone looking over my shoulder." The line exists; we just did not cross it.

Four design choices kept us on the right side:
\begin{itemize}[nosep, leftmargin=*]
    \item \textbf{Observation, not instruction.} "Your feed is 70\% food" vs.\ "You should diversify." The former respects autonomy; the latter prescribes.
    \item \textbf{Options, not actions.} The AI presents choices. Nothing changes until the user clicks. Dismissing the panel is always one tap away.
    \item \textbf{Gradual blending, not replacement.} Even after user confirmation, only 20--30\% of items change per cycle. No jarring full-page swaps.
    \item \textbf{Wait before speaking.} The AI initiates only after $\sim$20 items of scrolling. No pouncing.
\end{itemize}

We call this pattern \textit{gentle proactivity}: the AI takes initiative in starting the conversation, but the conversation is structured to maximize user agency at every decision point.

\subsection{Implications for Practitioners}

Three takeaways for anyone building recommendation products:

\textbf{1. Your feed already has the data to start the conversation.} The most well-received aspect of \system{} was the AI's initial insight---and that insight is computed from data every recommendation system already collects (category distributions, dwell times). The bottleneck is not data or models; it is the absence of an interface that turns observation into dialogue.

\textbf{2. Exploration needs an affordance, not just an algorithm.} Diversity re-ranking improves aggregate metrics, but users do not experience aggregate metrics. They experience a feed that feels repetitive. An explicit "let's explore" interaction---even a simple one---gives users agency over the process that algorithmic diversity cannot.

\textbf{3. Progressive commitment beats all-or-nothing.} The option $\rightarrow$ narrowing $\rightarrow$ blending pipeline lets users commit incrementally. Each step is low-risk, reversible, and informative. This encourages experimentation: a user who would never search for "outdoor photography" will happily click an option labeled "Surprise me."

\subsection{Relationship to InterQuest and Conversational Search}

\system{} and InterQuest~\cite{mei2025interquest} are complementary, not competing. InterQuest operates in a search context: the user has a query and is refining it through conversation. \system{} operates upstream: the user is scrolling a feed and has not formed any query. InterQuest's AI infers interests from search behavior; \system{}'s AI infers interests from passive browsing. InterQuest refines what the user is looking for; \system{} helps the user realize they are looking for something at all.

A natural product integration would hand off between the two: \system{} helps a user discover that they are interested in, say, outdoor photography. Once the user has that nascent interest, InterQuest's conversational search can help them refine it into "mirrorless cameras for landscape shooting under \$1000." The demand spectrum (Table~\ref{tab:spectrum}) is a pipeline, and different tools serve different stages.

\subsection{Limitations and Future Work}

\textbf{Curated content, not a live engine.} We used a fixed corpus of 320 items rather than a production recommendation system. This gave us experimental control (identical initial feeds) at the cost of ecological validity. Integrating \system{} with a live recommender and studying long-term effects is the obvious next step.

\textbf{Single session.} We measured first-time use. The novelty of AI-initiated interaction may inflate positive reactions. A longitudinal deployment is needed to test whether the effect habituates or sustains.

\textbf{LLM dependency.} The quality of \system{}'s insights depends on the LLM. Qwen3-VL-235B performed well in our study, but we have not tested generalization to smaller models or other content domains.

\textbf{One-size-fits-all proactivity.} Our formative study hinted that the optimal proactivity level is context-dependent (more during idle browsing, less during focused research). Adaptive proactivity calibration based on real-time engagement signals is a natural extension.

\textbf{Cultural specificity.} All participants were users of Chinese content platforms. Norms around AI proactivity and content browsing may differ elsewhere. Cross-cultural validation is needed.

\section{Conclusion}

Users know when their feed has gone stale, but they cannot tell you---or the search box---what they want instead. \system{} addresses this by flipping the interaction: the AI analyzes the feed, says what it sees, and offers low-effort options for exploring outward. In a mixed-design experiment ($n$=28), this AI-initiated approach nearly doubled exploration breadth, tripled the speed of discovering new content, and halved perceived interaction effort compared to a user-initiated chat with identical AI capabilities---without reducing users' sense of control. The mechanism is simple: the AI shifts the user from author ("What should I type?") to editor ("Yes, that one"), which turns out to be the difference between engagement and abandonment. We contribute a design pattern---\textit{gentle proactivity}---and argue that it applies wherever users experience vague dissatisfaction but lack the vocabulary to act on it.

\bibliographystyle{ACM-Reference-Format}
\bibliography{references}

\end{document}